\documentstyle[12pt]{article}

\newcommand{\al}{\alpha}
\newcommand{\ep}{\epsilon}
\newcommand{\de}{\delta}
\newcommand{\la}{\lambda}
\newcommand{\La}{\Lambda}

\newcommand{\deebar}{\bar{\partial}}

\newcommand{\df}{\stackrel{\rm def}{=}}

\newcommand{\lb}{\lbrack}
\newcommand{\rb}{\rbrack}

\newcommand{\msc}[1]{\mbox{\scriptsize #1}}
\newcommand{\dsp}{\displaystyle}

\newcommand{\br}{\mbox{{\bf R}}}
\newcommand{\bz}{\mbox{{\bf Z}}}

\newcommand{\bsz}{\msc{{\bf Z}}}
\newcommand{\bsr}{\msc{{\bf R}}}

\newcommand{\cA}{{\cal A}}
\newcommand{\cL}{{\cal L}}
\newcommand{\cG}{{\cal G}}
\newcommand{\cJ}{{\cal J}}
\newcommand{\cT}{{\cal T}}
\newcommand{\cO}{{\cal O}}
\newcommand{\cN}{{\cal N}}

\newcommand{\Ttop}{T^{\msc{top}}}

\newcommand{\Gtot}{G_{\msc{tot}}}

\newcommand{\ket}[1]{{|#1\rangle}}

\newcommand {\eqn}[1]{(\ref{#1})}

\newcommand{\cleqn}{\setcounter{equation}{0}}
\makeatletter
\@addtoreset{equation}{section}

\makeatother

\begin{document}
\vskip 7mm
\begin{titlepage}
 
 \renewcommand{\thefootnote}{\fnsymbol{footnote}}
 \font\csc=cmcsc10 scaled\magstep1
 {\baselineskip=14pt
 \rightline{
 \vbox{\hbox{hep-th/9909146}
       \hbox{UT-859}
       }}}

 \vfill
 \baselineskip=20pt
 \begin{center}
 \centerline{\Huge \rm Topological String on $AdS_3\times {\cal N}$}

 \vskip .8 truecm

 Yuji Sugawara\\
 {\sf sugawara@hep-th.phys.s.u-tokyo.ac.jp}

 \vskip .6 truecm
 {\baselineskip=15pt
 {\it Department of Physics,  Faculty of Science\\
  University of Tokyo\\
  Bunkyo-ku, Hongo 7-3-1, Tokyo 113-0033, Japan}
 }
 \vskip .4 truecm

 \end{center}

 \vfill
 \vskip 0.5 truecm

\begin{abstract}
\baselineskip 6.7mm

We study the topologically twisted string theory
on the general back-ground $AdS_3\times {\cal N}$
which is compatible with the world-sheet $N=2$ superconformal 
symmetry and is extensively discussed in the recent works \cite{GRBL}.
After summarizing the algebraic structure of the world-sheet
topological theory, we show that the space-time (boundary) 
conformal theory should be also topological.
We directly  construct the space-time topological conformal 
algebra (twisted $N=2$ superconformal algebra)
from the degrees of freedom in the world-sheet topological theory.
Firstly, we work on the world-sheet of the string propagating 
near boundary, in which we can safely make use of 
the Wakimoto free field representation. 
Secondly, we present a more rigid formulation of space-time 
topological conformal algebra  which is still valid far from the boundary
along the line of \cite{KS}.
We also discuss about the relation between this space-time topological 
theory and the twisted version of the space-time $N=2$ 
superconformal field theory given in \cite{GRBL}.

\end{abstract}

\setcounter{footnote}{0}
\renewcommand{\thefootnote}{\arabic{footnote}}
\end{titlepage}

\newpage
\baselineskip 7mm

\section{Introduction}

\cleqn
\hspace*{4.5mm}

String theory on $AdS \times {\cal N}$ back-ground 
has been getting great importance in the studies of 
$AdS/CFT$-duality \cite{Mal,GKP,Witten1,review}.
Among them, studies of string theory on  $AdS_3$ back-ground 
have old histories \cite{old,teschner}, in which  the several 
consistencies as the first quantized string theory 
have been intensively  discussed from the points of view of 
string propagating on a non-compact curved back-ground. 
More recently, a study emphasizing 
the relationship between the world-sheet picture  of
string theory and the description of boundary conformal field theory (CFT) 
on the  $AdS_3$-target \cite{BH,strominger} was 
initiated by Giveon-Kutasov-Seiberg \cite{GKS}, 
and many subsequent works which refine or extend it  
were carried out \cite{DORT}-\cite{Pet}. In particular  
the roles played by the short string sectors, which  
were lacking in \cite{GKS},  were investigated in \cite{DORT,KS}.

These approaches from the world-sheet picture of perturbative string  
have  opened up a new perspective to the studies of 
$AdS_3/CFT_2$-duality. Namely, we can consider quite general 
back-ground $AdS_3\times {\cal N}$, which {\em may not correspond
to any brane configuration\/}, as a consistent string vacuum.
Although such a back-ground is outside of original idea of  
Maldacena \cite{Mal},
which is based on brane theory, we can still expect the existence
of good holographic correspondence with some boundary 
CFT, because the asymptotic isometries  of 
$AdS_3$ considered by Brown-Hennaux \cite{BH}  should correspond 
to  infinite numbers of  conserved charges
in perturbative string theory.
In fact, in the framework given in  the recent works \cite{GRBL} 
(and their related works \cite{GiveonKP}),
we must generally assign a fractional value to
the level of $SL(2;\br)$ WZW describing the $AdS_3$-sector,
which corresponds to the brane charge in the usual setting of 
NS1/NS5.

In this paper we study the topological string on $AdS_3\times {\cal N}$
with the suitable $NS$ $B$-field,
whose world-sheet fermions have integral spins.
Although we again have no origin based on brane theory,
it is natural to expect that there exists 
some space-time (boundary) CFT which is dual to 
the world-sheet theory of topological string, because of the same reason
as above. We show this space-time CFT should be also topological,
and the main result of this paper is the construction of space-time
topological conformal algebra from the world-sheet point of view.
There are two non-trivial points in this construction:
The first is the difference of BRST charges between the topological 
string theory and the usual fermionic string theory. The second point is 
as follows: In the standard RNS formalism of fermionic string theory,
the space-time SUSY is realized by the spin fields, and so they are expected
to compose a part of the space-time superconformal algebra. 
However, in the case of topological string, 
we cannot consistently consider the  spin fields, 
since there do not exist the concepts such as  NS and R sectors.
This fact seems to make it  difficult  to construct the space-time topological
conformal algebra in the analogus form as that for the untwisted string 
theory. 

In order to construct the space-time conformal algebra
we first consider the near boundary approximation, in which
we can use  the Wakimoto free field realizations of 
affine $SL(2;\br)$ current algebra \cite{Wakimoto}.
We propose the  several conditions which should be satisfied by the correct 
space-time conformal algebra, and they lead us to a unique answer. 
Nextly, we work on a general string world-sheet propagating far from the
boundary according to the formalism developed in \cite{KS}. 
We present a candidate for the  space-time algebra in this framework. 
Although it has rather complicated form, we can prove 
without using the free field approximations
that it indeed generates the correct topological conformal algebra,
and it reduces to the previous result by the Wakimoto representation,
when taking the near boundary limit.

~

\section{Topological Twisting of 
Fermionic String on $AdS_3\times {\cal N}$}

\cleqn
\hspace*{4.5mm}

As was discussed in \cite{GRBL},
the world-sheet superconformal symmetry of 
the fermionic string theory on $AdS_3\times {\cal N}$
enhances to $N=2$,
if ${\cal N}$ has an affine $U(1)$-symmetry and ${\cal N}/U(1)$
defines an $N=2$ superconformal field theory (SCFT) on world-sheet.

To fix the notations, let us first present the field contents 
in this fermionic string theory. 
For more detailed arguments, see \cite{GRBL}.
\begin{itemize}
\item $AdS_3$-sector ~ $(j^A, \psi^A)$:
 
This sector has an affine $SL(2,\br)_k$ symmetry 
$\dsp J^A=j^A-\frac{i}{2}\ep^A_{~BC}\psi^B\psi^C$,
where the bosonic parts of currents $j^A$ generate
an affine $SL(2,\br)_{k+2}$ algebra and $\psi^A$ $(A=1,2,3)$
denote the free fermions in the adjoint representation.
\begin{eqnarray}
\psi^A(z)\psi^B(0) &\sim& \frac{\eta^{AB}}{z} ,~~~ \eta^{AB}\df 
\mbox{diag}(+,+,-) \\
j^A(z)j^B(0) &\sim& \frac{(k+2)\eta^{AB}}{2z^2}
+\frac{i\ep^{AB}_{~~~C}}{z}j^C(0)
   ,~~~(\ep^{123}=1 \mbox{ in our convention})\\
J^A(z)J^B(0) &\sim& \frac{k\eta^{AB}}{2z^2}
+\frac{i\ep^{AB}_{~~~C}}{z}J^C(0)
\end{eqnarray}
This sector has a world-sheet $N=1$ superconformal symmetry 
given by the next superconformal current;
\begin{equation}
G_{SL(2,\bsr)}
=\sqrt{\frac{2}{k}}\left(\psi^Aj_A -i\psi^1\psi^2\psi^3\right).
\end{equation}  
\item $U(1)$-sector ~ $(K,\chi)$:

We assume the existence of an affine $U(1)$ symmetry;
$\dsp K(z)K(0)\sim \frac{1}{z^2}$, and its fermionic partner;
$\dsp \chi(z)\chi(0) \sim \frac{1}{z}$. The $N=1$ superconformal current
in this sector is simply given by 
$G_{U(1)}= \chi K$.
\item ${\cal N}/U(1)$-sector:

We assume that this sector can be described by an $N=2$ $SCFT_2$

$\{T^{N=2}_{{\cal N}/U(1)},\, G^{\pm}_{{\cal N}/U(1)} ,\, 
J_{{\cal N}/U(1)}\}$.
\end{itemize}

Since we here consider a critical fermionic string, the total central
charge (except the ghost sector) should be equal to  $c=15$ 
($\dsp \hat{c}\equiv \frac{c}{3}=5$). Especially, 
the $N=2$ $SCFT_2$ for the ${\cal N}/U(1)$-sector 
must have the central charge $\dsp c=9-\frac{6}{k}$
($\dsp \hat{c}=3-\frac{2}{k}$), which implies 
\begin{equation}
J_{{\cal N}/U(1)}(z)J_{{\cal N}/U(1)}(0) \sim \frac{3-\frac{2}{k}}{z^2}.
\end{equation}

The $N=2$ structure is realized by considering 
the following formal decomposition of target space \cite{GRBL} (see also
the appendix B of \cite{GKS}); 
\begin{equation}
 \dsp AdS_3 \times {\cal N} \sim \frac{SL(2;\br)}{U(1)}\, \times \, U(1)^2 \,
\times \frac{{\cal N}}{U(1)} .
\label{decomp}
\end{equation}
We already assumed the ${\cal N}/U(1)$ sector has $N=2$ superconformal 
symmetry.  
We can describe the $SL(2;\br)/U(1)$ sector 
by the Kazama-Suzuki coset CFT \cite{KazS}, which is  essentially 
realized by the conformal fields $\{j^+,\, j^-,\, \psi^+,\, \psi^-\}$
($j^{\pm}\equiv j^1 \pm ij^2$, $\psi^{\pm}\equiv \psi^1\pm i\psi^2$), 
and the $U(1)^2$ sector is described  
by the standard ``Coulomb Gas'' 
representation of $N=2$ $SCFT$, whose members are two bosonic $U(1)$
currents $\{J^3,\, K\}$ and two fermions $\{\psi^3, \,\chi\}$.
It is convenient for our later discussions to combine these bosonic
currents and define the bosonic scalars  $\Phi^{\pm}$ as follows;
 \begin{equation}
  i\partial \Phi^{\pm} = 
 \frac{1}{\sqrt{2}} K \pm 
 \frac{1}{\sqrt{k}} J^3 ,
 \end{equation}
which have the OPE; 
$\dsp \partial \Phi^{\pm}(z)\partial \Phi^{\pm}(0)\sim 0$,
$\dsp \partial \Phi^{\pm}(z)\partial \Phi^{\mp}(0)\sim -\frac{1}{z^2}$.

Let us introduce the total $U(1)_R$ current defined by 
\begin{equation}
 \begin{array}{lll}
J_R&=&\dsp J_{SL(2;\bsr)/U(1)} +J_{U(1)^2}+J_{{\cal N}/U(1)} \\
&=&\dsp \left(\frac{1}{2}\psi^+\psi^- + \frac{2}{k}J^3\right) + \chi\psi^3
+J_{{\cal N}/U(1)}  .
\end{array}
\label{JR}
\end{equation}
According to the value of its charge,
the total $N=1$ superconformal current;
$\Gtot (z) = G_{SL(2;\bsr)} + G_{U(1)}+ 
(G^+_{{\cal N}/U(1)}+G^-_{{\cal N}/U(1)})$ is naturally
decomposed to the $N=2$ pieces; $\Gtot (z)=G_{N=2}^+(z)+G_{N=2}^-(z)$,
where
\begin{eqnarray}
G_{N=2}^{\pm}&=& G^{\pm}_{SL(2;\bsr)/U(1)}+G^{\pm}_{U(1)^2}
+G^{\pm}_{{\cal N}/U(1)} , \\
G^{\pm}_{SL(2;\bsr)/U(1)}&=& \frac{1}{\sqrt{k}}\psi^{\pm}j^{\mp}, 
  \label{N=2 G 1} \\
G^{\pm}_{U(1)^2}&=&\frac{1}{\sqrt{2}}(\chi\mp \psi^3)i\partial \Phi^{\pm},
  \label{N=2 G 2}
\end{eqnarray}
which defines an $N=2$ SCFT $\{ T^{N=2}, \, G_{N=2}^{\pm},\, J_R\}$.


It is a famous fact that we can construct a topological 
conformal field theory ($TCFT_2$) from an arbitrary 
$N=2$ $SCFT_2$ by the procedure 
of ``topological twisting'' \cite{Witten-top,EY}.
Roughly speaking, topological twisting means 
to shift the spins of the fermionic coordinates,
namely, to replace the physical fermion system with spin $1/2$
with the ghost system with spin $(0,1)$.
Usually topological twisting is defined with respect to 
the $U(1)_R$-current $J_R$ in  $N=2$ superconformal algebra (SCA) 
$\{T^{N=2},~ G_{N=2}^{\pm},~ J_R\}$ \cite{EY};
\begin{equation}
\Ttop_{EY}= T^{N=2}+\frac{1}{2}\partial J_R .
\label{EY twist}
\end{equation}
$\Ttop_{EY}$ indeed has central charge 0, since
\begin{equation}
J_R(z)J_R(0)\sim \frac{5}{z^2} \equiv \frac{\hat{c}}{z^2},
\end{equation}
and can be written in a BRST exact form 
$\dsp \Ttop_{EY}=\{Q_s , \, G_{N=2}^-\}$,
where $\dsp Q_s \df \oint dz \, G_{N=2}^+ $ denotes the BRST charge for TCFT.
However, this definition of twisting is not convenient for our purpose, 
because the fermionic coordinates
(and also the $\beta\gamma$-system in Wakimoto representation 
which we will introduce afterward) in the theory
have fractional spins with respect to the twisted stress tensor
$\Ttop_{EY}$ \eqn{EY twist}, and so the boundary conditions for these fields 
will become subtle. 
Therefore, we shall slightly modify the definition of topological
twisting. We consider the twisting  with respect to the ``total fermion
number current'' $J_f$\footnote
     {In fact, in the special case ${\cal N}= S^3\times T^4$,
      one will find that $J_f$ can be identified
      with $\dsp \frac{1}{2}\psi^+\psi^- + 
      \frac{1}{2}\chi^+\chi^- +\chi^3\psi^3 + 
      \la^{11}\la^{22} - \la^{12}\la^{21}$, where 
     $\{\chi^{\pm},\, \chi^3\}$ and $\la^{\al\dot{a}}$ $(\al, \dot{a}=1,2)$ 
     denote the fermionic
     coordinates along $S^3$ and $T^4$ respectively.} defined by
\begin{equation}
J_f \df \frac{1}{2}\psi^+\psi^- + \chi\psi^3 +
\left(J_{{\cal N}/U(1)} - \sqrt{\frac{2}{k}}K \right).
\label{Jf}
\end{equation}
The twisted stress tensor $\dsp \Ttop 
\df T^{N=2}+\frac{1}{2}\partial J_f$ also has zero central charge, 
because $\dsp J_f(z)J_f(0) \sim \frac{5}{z^2}$ holds, and 
as we will see below, $\Ttop$ can be  written  in  a 
$Q_s$-exact form. With respect to $\Ttop$,
$\dsp \xi_1\df \frac{1}{\sqrt{2}}\psi^+$, 
$\dsp \xi_2 \df \frac{1}{\sqrt{2}} (\chi -\psi^3)$ 
(and also $\gamma$ in Wakimoto rep.) 
have conformal weight 0 (we call them ``ghosts''), and 
$\dsp \eta_1\df \frac{1}{\sqrt{2}}\psi^-$, 
$\dsp \eta_2 \df \frac{1}{\sqrt{2}} (\chi +\psi^3)$ 
(and $\beta$ in Wakimoto rep.) 
have conformal weight 1 (``anti-ghosts'').
It may be worthwhile to notice that the ``fermion number
current'' for
the ${\cal N}/U(1)$-sector 
$\dsp J'_{{\cal N}/U(1)}\equiv 
J_{{\cal N}/U(1)} - \sqrt{\frac{2}{k}}K$
is in fact the same $U(1)$-current as the one made use of 
in order to define the spin fields in the papers \cite{GRBL} and has 
an integer level; 
$\dsp J'_{{\cal N}/U(1)}(z)J'_{{\cal N}/U(1)}(0) \sim \frac{3}{z^2}$.

The explicit form of our topologically twisted stress tensor is given as
follows;
\begin{equation}
\dsp \Ttop = \frac{1}{k}j^Aj_A +\frac{1}{2}K^2 - \frac{1}{\sqrt{2k}}\partial K
-\eta_1\partial \xi_1 -\eta_2\partial \xi_2 
+\left(T^{N=2}_{{\cal N}/U(1)}+\frac{1}{2}\partial J_{{\cal N}/U(1)}\right).
\label{top total T} 
\end{equation}
This topological theory can be naturally decomposed to 
the three independent sectors mentioned above;
\begin{equation}
\Ttop=T_1+T_2+T_{{\cal N}/U(1)}.
\end{equation}
Each of them is described by the ``topological conformal algebra'' (TCA):
$\{T,\,G^{\pm},\,J\}$,
which is defined as the twisted $N=2$ superconformal algebra and 
our convention for this algebra is presented in appendix A.  
The generators belonging to 
the different sectors (anti-)commute with one another.
This is the  structure analogous to that for the topologically
twisted gauged WZW model \cite{NS}. 
We summarize this aspect as follows;
\begin{enumerate}
\item $SL(2,\br)/U(1)$-sector (twisted Kazama-Suzuki for 
$SL(2,\br)_{k+2}/U(1)$ with $\dsp \hat{c}=1+\frac{2}{k} $):
 \begin{equation}
 \left\{
 \begin{array}{lll}
 T_1&= &\dsp \frac{1}{k}(j^Aj_A+J^3J^3) - \eta_1\partial \xi_1 + 
 \frac{1}{k}\partial J^3  \\
 G_1^+ &=& \dsp \frac{1}{\sqrt{k}}\xi_1 j^- \\
 G_1^- & = & \dsp \frac{1}{\sqrt{k}}\eta_1 j^+ \\
 J_1 & =& \dsp  \xi_1\eta_1 + \frac{2}{k}J^3 
 \end{array}
 \right.
 \end{equation}
$$ (J^3 \equiv j^3-\eta_1\xi_1) $$
\item $U(1)^2$-sector (twisted $N=2$ Coulomb-Gas representation,  $\hat{c}=1$)
 \begin{equation}
 \left\{
 \begin{array}{lll}
 T_2 &= &\dsp -\partial \Phi^+ \partial \Phi^- - \eta_2\partial \xi_2
 - \frac{1}{\sqrt{k}}i\partial^2 \Phi^+ \\
 G^+_2 &=& i\xi_2 \partial \Phi^+ \\
 G^-_2 &=&\dsp  i\eta_2 \partial \Phi^- -\frac{1}{\sqrt{k}}\partial \eta_2 \\
 J_2 &= &\dsp \xi_2\eta_2-\frac{1}{\sqrt{k}}i\partial \Phi^+
 \end{array}
 \right.
 \end{equation}
\item ${\cal N}/U(1)$-sector (twisted in the usual sense of \cite{EY},  
 $\dsp \hat{c}=3-\frac{2}{k}$) : 

$\dsp \{T_{{\cal N}/U(1)}
(\equiv T^{N=2}_{{\cal N}/U(1)}+\frac{1}{2}\partial J_{{\cal N}/U(1)}), 
 G^{\pm}_{{\cal N}/U(1)} , J_{{\cal N}/U(1)}\}$.
\end{enumerate}

By our construction it is obvious that $G^{\pm}_1$ are equal to 
$G^{\pm}_{SL(2;\bsr)/U(1)}$ defined previously \eqn{N=2 G 1}.
However, one should remark that $G^-_2\neq G^-_{U(1)^2}$,
although $G^+_2= G^+_{U(1)^2}$ still holds. (Namely, 
$\Gtot^+\equiv G^+_1+G^+_2 + G^+_{{\cal N}/U(1)}$ is equal to
$G_{N=2}^+$, but $\Gtot^-\neq G_{N=2}^-$.)
Generally, for an arbitrary parameter $\al$,
 \begin{equation}
 \left\{
 \begin{array}{lll}
 T_2 &= &\dsp -\partial \Phi^+ \partial \Phi^- - \eta_2\partial \xi_2
 - \frac{\al}{\sqrt{k}}i\partial^2 \Phi^+ \\
 G^+_2 &=& i\xi_2 \partial \Phi^+ \\
 G^-_2 &=&\dsp  i\eta_2 \partial \Phi^- -\frac{\al}{\sqrt{k}}\partial \eta_2 \\
 J_2 &= &\dsp \xi_2\eta_2-\frac{\al}{\sqrt{k}}i\partial \Phi^+
 \end{array}
 \right.
\end{equation}
compose the TCA with $\hat{c}=1$.
The case of $\al=0$ corresponds to the topological twisting of
\cite{EY} \eqn{EY twist} and we have $G^-_2=G^-_{U(1)^2}$. 
The twist by $J_f$ \eqn{Jf} corresponds 
to $\al=1$, in which we have $G^-_2\neq G^-_{U(1)^2}$ as 
we remarked above.

We must also consider the ghost system associated with
the gauge fixing of diffeomorphism. In the fermionic string
theory before twisting, we have the spin $(2,-1)$ $bc$-ghosts 
and the  spin $(3/2,-1/2)$ $\beta\gamma$-ghosts, which yield an $N=1$
$SCFT_2$ with $c=-15$ that cancels the central charge of ``matter sector''.
On the contrary,  
because  the matter sector in the topological string theory 
has already $c=0$, the ghost sector should become a $c=0$ CFT, too.
This leads us to  the system of fermionic ghosts with spin $(2,-1)$
and bosonic ghosts also with spin $(2,-1)$.  It is no other than 
the ``topological gravity sector'' \cite{TG}, which also has a sturcture of 
twisted $N=2$ with $\hat{c}=-3$.
Through this paper we shall use the symbols $\beta$, $\gamma$ for
the Wakimoto free fields, and 
describe the topological gravity sector 
by the standard bosonizations \cite{FMS} to avoid confusion, 
that is, we set $b=e^{-i\sigma}$, $c=e^{i\sigma}$, 
$\beta=\partial \xi e^{-\rho}$, $\gamma=\eta e^{\rho}$, where
$(\xi, \eta)$ is a spin (0,1) fermionic fields, and $\sigma$, $\rho$
are scalar fields with the back-ground charges\footnote
    {In our convention the back-ground charge $Q$ is defined by 
    $c=1+3Q^2$.}$Q_{\sigma}=3i$, $Q_{\rho}=-3$, which have the OPEs;
$\dsp \sigma(z)\sigma(0)\sim -\log z$, $\rho(z)\rho(0) \sim -\log z$.
It is also convenient to define  $\phi \df -\rho + i\sigma$.
The TCA in this sector is generated by the following currents;
\begin{equation}
\left\{
\begin{array}{lll}
T_{TG} &= &\dsp -\frac{1}{2}\partial \phi \partial \phi^* 
+ \frac{3}{2}\partial^2 \phi -\eta \partial \xi \\
G^+_{TG} &=& \eta e^{-\phi} \\
G^-_{TG} &=& \left\{-\partial^2 \xi +\partial \xi
 (\partial \rho -2i\partial \sigma)\right\}e^{\phi} \\
J_{TG} &=& -2\partial \rho + i\partial \sigma
\end{array}
\right. 
\label{TG}
\end{equation}
The BRST charge\footnote{Strictly speaking, 
     the precise  BRST charge  should be $Q_{BRST}\df Q_s
   +Q_{\msc{diff}}$, where $Q_{\msc{diff}}$ is defined by 
   $$
   Q_{\msc{diff}} =\oint \left\{c\left(\Ttop+\frac{1}{2}T_{TG}\right)
   - \gamma\left(\Gtot^- + \frac{1}{2}G^-_{TG} \right) 
   \right\}.   
   $$ 
    But we can easily find that $\dsp UQ_s U^{-1} = Q_{BRST}$ holds,
    where we set $U = e^{\oint c \left(\Gtot^- + \frac{1}{2}G^-_{TG}\right)}$,
    and hence we can consistently regard $Q_s$ as the BRST charge of 
    the theory.}
for our topological theory is defined by
\begin{equation}
Q_s \df \oint (G^+_1+G^+_2+G^+_{{\cal N}/U(1)} +G^+_{TG}) 
 (\equiv \oint (\Gtot^+ +G^+_{TG}) ) .
\label{BRST}
\end{equation} 
The stress tensor in each sector $T_i$ can be written in the BRST exact
form;
\begin{equation}
 T_i=\{Q_s,~ G^-_i \},
\end{equation}
and so does the total stress tensor $\Ttop$ \eqn{top total T}. 

The physical states are defined by
\begin{equation}
 (Q_s+\bar{Q}_s)\ket{\mbox{phys}}=0 , 
\label{phys}
\end{equation}
according to the generalities of TCFT.  
As is well-known, there are natural two choices for the chiralities of 
$\bar{Q}_s$ (``topological A-model'' or ``topological B-model''). However, 
the detailed discussions for this subject is beyond the scope of this paper.
We shall only focus on the left (or right) mover 
in the following discussions, and consider the physical condition
\eqn{phys} simply as
\begin{equation}
 Q_s\ket{\mbox{phys}}=0 , ~~~ \ket{\mbox{phys}}\sim \ket{\mbox{phys}}
                                  +Q_s\ket{\mbox{any}}.
\label{phys2} 
\end{equation}
The physical states correspond to the chiral primary states in the
untwisted $N=2$ theory on world-sheet.
Moreover, since 
\begin{equation}
J^3_0+\sqrt{\frac{k}{2}}K_0 
\left(\, \equiv \oint (j^3 + \xi_1\eta_1 + \sqrt{\frac{k}{2}}K) \, \right)
 = \left\{ \, Q_s, \, \sqrt{k}\oint \eta_2 \, \right\}  
\end{equation}
holds, arbitrary physical states must satisfy 
\begin{equation}
\left(J^3_0+\sqrt{\frac{k}{2}}K_0\right)\ket{\mbox{phys}}=0 
~~~(\mbox{mod BRST}) ,
\label{space-time BPS}
\end{equation}
which is no other than the BPS condition in space-time.
It implies that the physical states are also in one-to-one
correspondence to the 1/2 BPS states, in other words
the chiral primary states in  the space-time theory derived 
from the untwisted world-sheet theory. 

From this observation we can immediately conclude that the space-time CFT
associated to our topological string must be also topological.
In fact, since we now work on the $AdS_3$ back-ground,
it is natural to expect that 
the $Q_s$-invariant  Virasoro operators $\{\cL_n\}_{n\in \bsz}$ 
of Brown-Hennaux exist, and if so, they must satisfy
\begin{equation}
 [J^3_0 ,~ \cL_n] = n\cL_n, ~~~[K_0,~\cL_n]=0 ,
\end{equation}
because of the requirement of $SL(2;\br)$-symmetry on $AdS_3$.
However, this means that $\cL_n \ket{\mbox{phys}}$ cannot satisfy
the condition \eqn{space-time BPS} for  any
physical state $\ket{\mbox{phys}}$, 
{\em unless all of the $\{\cL_n\}$ are  BRST-exact.}
In this way, we have found that 
topological string on $AdS_3$ back-ground should lead to 
the space-time conformal symmetry with the Virasoro operators
having the BRST-exact forms, namely, a topological conformal field
theory. 

At this stage a few natural questions may arise;
Does this space-time TCFT have a twisted $N=2$ algebra
describing its local symmetry in the same way as the familar models of 
TCFT in two dimension?  Moreover, if it exists, 
can we regard it as the twisted version of the space-time $N=2$ algebra
constructed in \cite{GRBL}?  

The answer for the first question is  {\em Yes\/}.
In the next section we will construct 
the suitable space-time conformal algebra
from the world-sheet field contents.  
At first glance, the answer for the second question seems also {\em Yes\/},
because the most natural choice of
$\cL_0$-operator is as follows (up to normalizaiton, of course);  
\begin{equation}
\cL_0 = - \left(J^3_0 +\sqrt{\frac{k}{2}} K_0\right),
\end{equation}
which is a BRST-exact operator as seen above.
This is indeed the $\cL_0$-operator in 
the twisted version of space-time $N=2$ theory in \cite{GRBL},
since $\cL_0^{N=2}=-J^3_0$, and the $U(1)_R$ charge operator 
is $\cJ^{N=2}_0 =\sqrt{2k}K_0$.
However, this is not the whole story.
We must also care about the  detailed structure 
of higher Virasoro generators to obtain this answer completely.      
In the next section we will discuss this point.

~


\section{Topological Conformal Symmetry in Space-time}

\cleqn
\hspace*{4.5mm}

As we already demonstrated, the space-time conformal theory for
the topologically twisted string  should also become  
topological. In this section we construct the TCA in space-time.
We first work in the near boundary region,
where the string coordinates on $AdS_3$ become free fields, 
and next present the more rigid framework that is still valid
far from the boundary.

\subsection{Space-time Topological Conformal Algebra 
in the Near Boundary Approximation}

First of all, let us remember the (bosonic) $AdS_3$ sector is 
described in quantum level  by the following Lagrangian \cite{GKS};
\begin{equation}
\cL=\partial\varphi\deebar\varphi- \sqrt{\frac{2}{k}}R^{(2)}\varphi
+\beta\deebar\gamma+\bar{\beta}\partial\bar{\gamma}
-\beta\bar{\beta}\exp \left(-\sqrt{\frac{2}{k}} \varphi\right),
\label{ads3 lagrangian}
\end{equation} 
where $R^{(2)}$ means the curvature on world-sheet.
As was clarified in \cite{DORT}, in the near boundary region 
($\varphi \sim +\infty$), since  the interaction term 
$\dsp \sim \beta\bar{\beta}\exp \left(-\sqrt{\frac{2}{k}} \varphi\right)$ 
is suppressed, we can safely make use of
the Wakimoto free field representations \cite{Wakimoto}
for the $AdS_3$-sector.
The (bosonic) $SL(2;\br)$-current can be realized as follows; 
\begin{equation}
\left\{
\begin{array}{l}
\dsp  j^3 = \beta\gamma + \sqrt{\frac{k}{2}}\partial \varphi \\
      j^- = \beta \\
      j^+ = \beta \gamma^2+\sqrt{2k}\gamma\partial \varphi 
      +(k+2)\partial\gamma 
\end{array}
\right.
\end{equation}
where $\dsp \beta(z)\gamma(0)\sim \frac{1}{z}$, 
$\varphi(z)\varphi(0)\sim -\log z$ in our convention.
The stress tensor in this sector is given by
\begin{equation}
\dsp  T(z)= -\frac{1}{2}(\partial \varphi)^2 - 
\frac{1}{\sqrt{2k}}\partial^2 \varphi + \beta \partial \gamma . 
\end{equation}

It is important to notice that the interaction term  
$\dsp \sim \beta\bar{\beta}\exp \left(-\sqrt{\frac{2}{k}} \varphi\right)$ 
is so-called ``screening charge'',  which does not break
at least perturbatively the conformal invariance as well as the 
affine $SL(2;\br)$ symmetry  on the world-sheet.
This implies that the results given in the previous section 
are still valid in a generic region with $\varphi \sim \mbox{finite}$,
in which the free field approximations to the coordinates
$\beta$, $\gamma$, $\varphi$ are  not reliable. In fact, we did not 
need the explicit realizations of $SL(2;\br)$-currents for the
discussions in the previous section.

However, since the Wakimoto rep. is very powerful
for the purpose in this section, which is  to construct
the suitable space-time conformal algebra, we shall first work in
the near boundary region $\varphi \sim +\infty$. 
In the next subsection we will  consider the general region according to 
the framework developed in \cite{KS}.

The goal in this subsection is to construct 
the full space-time conformal algebra from the degrees of freedom 
in the world-sheet topological theory
in the analogous form as that given in \cite{GKS}. 
The first guess for the answer may be the twisted version of 
the space-time $N=2$ SCA for the untwisted string \cite{GRBL}
as mentioned in the previous section. 
Namely, one might suppose that  
\begin{equation}
\cL_n = \cL_n^{N=2}
-\frac{1}{2}(n+1)\cJ^{N=2}_n,
\label{first guess}
\end{equation}
where $\cL_n^{N=2}$, $\cJ^{N=2}_n$ 
are the stress tensor and the $U(1)_R$ current given in \cite{GRBL}.
Unfortunately, although this expectation seems to be natural, 
it is {\em not\/} correct. 
This is  because the R.H.S of \eqn{first guess} is {\em not\/}
BRST-invariant {\em in the sense of topological string theory\/}
(except $n=0$).
Furthermore we have another  serious difficulty: In the untwisted theory
the space-time supercurrents are made up of the spin fields. 
But, in the topological string theory, we have only the fermionic
fields with integral spins (ghost system) on world-sheet, and hence 
only the periodic boundary condition is  allowed. In this case 
we cannot consider the spin fields, which intertwine the NS-sector 
with the R-sector in the usual fermionic string theory. 
Therefore we must look for the candidates of space-time conformal 
currents only within the ordinary vertex operators 
(which has no pieces of the spin fields).

We here assume  the conditions which the space-time conformal
algebra $\{\cG^{\pm}_n,~ \cL_n ,~ \cJ_n\}$
in our topological model should satisfy;
\begin{enumerate}
\item 
They should generate a topological conformal algebra 
(twisted $N=2$ superconformal algebra).
Especially, $\{\cL_n \}$ generates the Virasoro algebra with zero 
central charge (irrespective of the value 
of $\dsp \oint \gamma^{-1}d\gamma$).
\item $\{ \cL_n \}$ act on the (bosonic) boundary coordinates as 
suitable local conformal transformations  in the same way as those given
in \cite{GKS}.
\item $\{\cL_n \}$ and $\{\cG_n^+\}$ should be $Q_s$-invariant.
\item $\dsp \cG^+_0 = Q_s \equiv \oint (\Gtot^+ + G^+_{TG}) $ 
holds (up to normalization).
\item All of the  generators can be written in the form;
$\dsp \cA_n=\oint dz \, a_n(z)$, where $a_n(z)$ is a spin 1 primary
field on the world-sheet, that is, 
it is a spin 1 current {\em with no back-ground charge
on the  world-sheet};
$$ 
\Ttop(z)a_n(w) \sim 0\times \frac{1}{(z-w)^3}+\cdots  .
$$
\end{enumerate}
The first and second conditions are quite natural.
But there is one important remark; We require the first condition
is satisfied ``strictly'', that is, {\em not only up to $Q_s$-exact terms\/}.
Otherwise, we cannot determine the space-time conformal algebra at all, 
because the Virasoro generators themselves are now  $Q_s$-exact and hence
the algebra becomes trivial identities  in the  sense of up to
$Q_s$-exact terms.

One might ask about the third condition: Why do not we impose 
the $Q_s$-invariance on $\cG^-$ and $\cJ$, too?
However, it is not peculiar in topological conformal field theory.
Usually,
$\{Q_s ,~\cG^-\}$ is not zero, rather equal to the stress tensor,
and the ``ghost number current'' $\cJ$ does not also commute with $Q_s$.
As was already demonstrated, the assumption of the $Q_s$-invariance
of $\cL_n$ inevitably leads to the $Q_s$-exactness of it. 
This implies the topological invariance in the space-time theory.

The fourth condition means that the BRST-charge in the space-time
topological theory is equal to the one for the world-sheet theory.
This assumption might be too strong. 
But, from the viewpoint of $AdS/CFT$-duality,
all of the physical observables in the world-sheet theory
are expected to become  also physical in the sense of space-time.
This is the simplest assumption to realize this correspondence.

The fifth condition is more subtle. In the usual (untwisted) string theory,  
this is nothing but a result of BRST-invariance. However, we cannot
derive this condition from the $Q_s$-invariance.
Nevertheless the reason why we believe this condition is natural is as
follows: In a general $CFT_2$ the back-ground charge term 
$\dsp T(z)a(w)\sim -\frac{q}{(z-w)^3} \cdots$ is related with
the anomaly term on {\em a curved world-sheet\/}; 
$\deebar a(z) \sim q R^{(2)}$.  
So, if we require that all of 
the generators in the space-time conformal algebra  
should be  conserved charges {\em even on the curved world-sheet\/},
we need impose this condition on it. 
This condition is quite important as a guiding principle 
to look for the correct answer. 
For example, the space-time $U(1)_R$ current in the untwisted model
has a non-zero back-ground charge with respect to the stress tensor
$\Ttop$ \eqn{top total T} (after the replacement
$\psi^+ \rightarrow \sqrt{2} \xi_1$, 
$\psi^- \rightarrow \sqrt{2} \eta_1$, etc.), 
and hence we cannot take this current as the candidate for
the ``ghost number current'' $\cJ_n$ in the space-time TCA.

These five conditions can lead us to the unique answer
(up to normalizations). 
Let us exhibit this result.   
We here  use the abbreviated notation 
$\dsp \oint \equiv \oint\,\frac{dz}{2\pi i}$;
\begin{eqnarray}
\cG_n^+ &=& \left\lb 
Q_s, \, -\sqrt{k}\oint \gamma^n \hat{J} \right\rb \nonumber \\
 &=& \oint \left\{\sqrt{k}\gamma^n(\Gtot^+ +G^+_{TG}) 
-n(\gamma^{n-1}\xi_1+\gamma^n\xi_2)\hat{J} 
- (k+1)\gamma^n \partial \xi_2\right\} \nonumber  \\
\cG_n^- &=& \oint 
\left\{n\eta_1 \gamma^{n+1} -(n+1) \eta_2\gamma^n \right\} \nonumber \\
\cL_n &=& \{ Q_s ,\, \sqrt{k}\cG^-_n \} \nonumber \\
  &=& \oint \left\{n(j^-+\eta_1\xi_2)\gamma^{n+1}
-(n+1)(j^3-\eta_1\xi_1+\sqrt{\frac{k}{2}}K)\gamma^n 
-n(n+1)\gamma^n R\right\} \nonumber \\
\cJ_n &=& \oint \gamma^n\hat{J} 
+ \left\{Q_s,\, \sqrt{k}\oint \gamma^n(-\gamma\eta_1+\eta_2)\right\} 
  \nonumber \\
&=& \oint \gamma^n(J_f -\partial \rho + \sqrt{2k}K + nR) 
\label{stTCA1}
\end{eqnarray}
where we set 
\begin{eqnarray}
\hat{J} &=& J_f -\partial \rho + \sqrt{\frac{k}{2}}K -
 \sqrt{\frac{k}{2}} \partial \varphi , \\
R&=& \eta_1\xi_1+\gamma\eta_1\xi_2-\gamma^{-1}\eta_2\xi_1
-\eta_2\xi_2 .
\end{eqnarray}
The integrands of these operators  are actually spin 1 conformal 
fields with no back-ground charges 
with respect to $\Ttop$ \eqn{top total T}. 
After some straightforward and lengthy calculations, we can show 
they generate a TCA with $\dsp \hat{c}=2k\oint\,\gamma^{-1}d\gamma$.
This $\hat{c}$ is equal to that for the space-time theory for 
the untwisted string ($\dsp c^{N=2} = 6k\oint \gamma^{-1}d\gamma$),
although our space-time TCA \eqn{stTCA1} has a form  which is quite
different from the twisted version of space-time $N=2$ SCFT. 
In fact,  $\dsp \cL_{0}=- \left(J^3_0 +\sqrt{\frac{k}{2}}K_0 \right)$ holds
as we mentioned in the previous section, but in general
$\dsp \cL_n \neq \cL_n^{N=2} -\frac{1}{2}(n+1)\cJ_n^{N=2} $. 

We here make one comment: 
The topological gravity sector (especially, the scalar field
$\partial \rho$) seems to play an important  role
in order to obtain  the correct conformal algebra in space-time.  
In fact, it is not so difficult to show that  
we cannot construct the conformal algebra 
which satisfies all the five conditions above without topological
gravity sector. 
Why it is so may be interesting problem.
We would like to further discuss this point (from a geometrical point
of view, maybe) elsewhere.

Before leaving the arguments in the near boundary approximation,
let us discuss the ``space-time vacuum'' (long string vacuum)
which realizes $\dsp \oint \gamma^{-1}d\gamma=p\neq 0$ in topological 
theory. By making use of the familiar bosonizations: 
$\beta=i\partial Y e^{-X-iY}$, $\gamma=e^{X+iY}$, where
$ X(z)X(0)\sim -\ln z$, $Y(z)Y(0)\sim -\ln z$, 
we can immediately  make up a $Q_s$-invariant (and not $Q_s$-exact) 
state having such a property within the large Hilbert space of
$\beta\gamma$\footnote
  {It may be worthwhile to emphasize
   that we must here work on the large Hilbert space
  rather than the reduced space  projected by $\dsp \oint e^{iY} =0$. 
  This is because we need 
  the {\em negative\/}  powers of $\gamma$ as well as the positive powers 
  in the original construction of the space-time Virasoro algebra by 
  GKS \cite{GKS}.}. We set
\begin{equation}
 \begin{array}{lll}
 V &\df& \xi_1e^{-X}e^{\phi}\equiv \xi_1ce^{-X}e^{-\rho} \\
 V^{-1}&\df& \eta_1 e^X e^{-\phi}\equiv \eta_1be^Xe^{\rho} .
\end{array}
\label{V1}
\end{equation}
They are $Q_s$-invariant: $\lb Q_s,~ V^{\pm 1}\rb=0$ and 
$V\cdot V^{-1}=V^{-1}\cdot V=1$. Moreover they satisfy
\begin{equation}
\left\lb \oint\, \gamma^{-1}d\gamma ,~ V^{\pm 1}\right\rb =\pm V^{\pm 1}.
\end{equation}
It is easy to see we can consider the operator products $V\cdot V\cdots$, 
$V^{-1}\cdot V^{-1}\cdots$ without any regularization of divergenece.
Therefore we can introduce the following space-time vacuum in our
topological model;
\begin{equation}
\ket{\mbox{vac};p}\df V^p\ket{0} ~~~ (p\in \bz) ,
\label{vac1}
\end{equation}
which is manifestly $Q_s$-invariant and satisfies
\begin{equation}
 \oint \,\gamma^{-1}d\gamma\ket{\mbox{vac};p} =p\ket{\mbox{vac};p} 
\end{equation}
for any winding number $p\in \bz$.

It is interesting to compare this result with that for the untwisted
theory. The operator $V_{\msc{untwisted}}$ 
in the untwisted fermionic string theory should 
be given by
\begin{equation}
 V_{\msc{untwisted}} \df  \psi^+ ce^{-X}e^{-\rho}.
\label{V2}
\end{equation}
This operator is BRST-invariant (and not exact)
in the framework of the untwisted fermionic string, and also 
satisfies $[\cL^{GKS}_n ,~ V]=0$ (mod BRST) for $\forall n \geq -1$,
where $\cL^{GKS}_n$ denote  the space-time Virasoro generators given in
\cite{GKS}.
The space-time vacuum for $p=1$ is defined as 
\begin{equation}
\ket{\mbox{vac}}_{\msc{untwisted}} \df V_{\msc{untwisted}}\ket{0} ,
\label{vac2}
\end{equation}
which is no other than 
the $(-1)$-picture version of that introduced in \cite{HS}.
(Remember that in this paper 
we write the bosonized  superghost as ``$\rho$'' instead of the usual
notation ``$\phi$''.).

Since one can again define the operator products 
$V_{\msc{untwisted}}\cdot V_{\msc{untwisted}} \cdots$ without any
regularization, one might imagine that the space-time vacuum for any
$p\in \bz$ can be defined simlarly as \eqn{vac1}.
But this naive guess  does {\em not\/} work. In fact, we can prove that 
$V_{\msc{untwisted}}^p$ $(\forall p\geq 2)$ becomes BRST-exact,
and moreover the operator corresponding to $V^{-1}$ (which is given
by replacing $\eta_1$ with $\psi^-$ in the expression \eqn{V1}) 
does not become BRST-invariant in the untwisted theory. 
This observation suggests the claim given in \cite{HS}: 
One can only construct the space-time vacuum for $p=1$
on the single string world-sheet, and in order to realize
the cases $p\geq 2$ one should consider the multi-string system
desribed by $Sym^p$ orbifold theory. This claim is consistent with
the naive estimation of physical degrees of freedom, because
the winding number $p$ should correspond to a brane charge
(NS1 charge), and is related with the upper bound for the  $U(1)_R$
charges of the chiral primary fields in the boundary CFT \cite{MS}.  

On the other hand, as seen above,  in the topological
string we can make up the space-time vacuum for an {\em arbitrary\/} winding
$p$ {\em on a single string world-sheet\/}. It is not a contradiction,
since the winding in topological string does not correspond to any
brane charge, and we can show that we have no upper bound for the
$U(1)_R$ charge related to the value of $p$ in the topological model. 
The topological gravity sector seems to play an essential role 
in this feature  because of the existence  
of ``gravitational descentants''.
Of course, in order to complete this discussion 
we need to analyse carefully the detailed structure of physical spectum 
in the topological theory. This analysis will be given in \cite{sugawara}.

~

\subsection{Space-time Topological Conformal Algebra in Finite $\varphi$}

Now let us consider the  more general situation when  the string
world-sheet exists far from boundary. In this case the free field 
description does not work, 
although we still have an affine $SL(2;\br)$ symmetry.
To this aim it is useful to introduce auxiliary parameters  $x$, $\bar{x}$, 
which are naturally identified with the coordinates on
boundary, according to the works \cite{teschner,DORT,KS,BDM},
and to write down directly the space-time conformal
currents as $\cT(x)$, $\cG^{\pm}(x)$, $\cJ(x)$ instead of 
dealing with their Fourier modes. 
In particular, we shall follow the convention of \cite{KS}, and refer 
to the above papers \cite{teschner,DORT,KS,BDM}
for the more detailed arguments.

First of all, let us introduce
\begin{equation}
j(x;z) \df  2xj^3(z) -x^2j^-(z)-j^+(z) ,
\label{jx}
\end{equation}
which satisfies the following OPE;
\begin{equation}
j(x;z)j(y;w) \sim (k+2)\frac{(y-x)^2}{(z-w)^2} +\frac{1}{z-w}
 \left\{(y-x)^2\partial_y -2(y-x)\right\}j(y;w) .
\label{jj}
\end{equation}
The counterpart  in the right mover $\bar{j}(\bar{x};\bar{z})$
is similarly defined.
It is also convenient to  introduce the notation 
\begin{equation}
\tilde{j}(x;z) \df \dsp j(x;z)+2x\sqrt{\frac{k}{2}}K(z) . 
\label{j tilde x} 
\end{equation}
These currents \eqn{jx}, \eqn{j tilde x} 
have the weight (1,0) for world-sheet and 
the weight $(-1,0)$ for space-time. (The boundary coordinates $x$, $\bar{x}$
should have  the space-time dimensions $(-1,0)$, $(0,-1)$ respectively).
The primary field in space-time is defined by
\begin{equation}
   \Phi_h(x,\bar{x};z,\bar{z}) \df 
 \frac{1}{\pi}\left(\frac{1}{|x-\gamma(z)|^2
 e^{\frac{\varphi(z)}{\sqrt{2k}}} +  e^{-\frac{\varphi(z)}{\sqrt{2k}}}  }
 \right)^{2h} ,
\label{Phi h}
\end{equation} 
which is an analog of  the bulk-boundary Green function.
This operator should have the following OPE with the current $j(x;z)$;
\begin{equation}
j(x;z)\Phi_h(y,\bar{y};w,\bar{w}) \sim 
\frac{1}{z-w}\left\{(y-x)^2\partial_y+2h(y-x) \right\}
                          \Phi_h(y,\bar{y};w,\bar{w}),
\end{equation}
which means $\Phi_h$ indeed has the space-time conformal weight $(h, h)$.
For our purpose the $h=1$ case is the most important;
\begin{equation}
j(x;z)\Phi_1(y,\bar{y};w,\bar{w}) \sim 
\frac{1}{z-w}\partial_y \left\{(y-x)^2 \Phi_1(y,\bar{y};w,\bar{w})\right\},
\label{Phi 1 OPE}
\end{equation}
and it is easy to see 
\begin{equation}
\lim_{\varphi\rightarrow +\infty} \Phi_1(x,\bar{x};z,\bar{z})
=\de^2(x-\gamma(z)).
\end{equation}

We must further prepare the fermionic partners of the above
currents. We set
\begin{equation}
\begin{array}{lll}
\xi(x;z)&\df& \xi_1(z)+x\xi_2(z)\\
\eta(x;z)&\df&-x\eta_1(z)+\eta_2(z),
\end{array}
\end{equation}
which satisfy the OPE;
\begin{equation}
\xi(x;z)\eta(y;w) \sim -\eta(x;z)\xi(y;w) \sim \frac{x-y}{z-w}.
\label{xi eta}
\end{equation}
$\xi(x;z)$ should have the world-sheet weight $(0,0)$ and the space-time 
weight $(-1,0)$. $\eta(x;z)$ should have $(1,0)$ for world-sheet and
$(0,0)$ for space-time.

Under these preparations we can propose 
the following currents as the suitable topological conformal
algebra  in space-time $\{\cG^{\pm}(x),\, \cT(x),\, \cJ(x)\,\}$.
To avoid the complexity of notations
we shall use the abbreviations;
$\xi\equiv \xi(x;z)$, $\eta\equiv \eta(x;z)$, 
$j\equiv j(x;z)$ etc.;
\begin{equation}
\begin{array}{lll}
  \cG^+(x)&=& \lb Q_s ,~-\sqrt{k}\cJ(x)\rb  \\ 
&=&  \dsp -\frac{1}{k+2}
 \int d^2z  \,\left\lb (\bar{j}\Phi_1) \{\sqrt{k}(G^+_{\cN/U(1)}+G^+_{TG})
 -\partial^2_x \tilde{j} \xi 
 -(k+1)\partial_z\partial_x\xi \}  \right. \\
 & & \hspace{1in} \dsp \left. 
  + \partial_x(\xi\bar{j}\Phi_1)\left(J_{\cN/U(1)}-
 \partial \rho+ (k-1)\sqrt{\frac{2}{k}}K +2\partial_x\eta\xi\right)
    \right. \\
 & & \hspace{1in} \dsp \left.  - \partial_x(\bar{j}\Phi_1) 
 ( \frac{1}{2}\partial_x \tilde{j}\xi +\eta\partial_x \xi \xi+ \partial_z \xi
  )\right\rb \\
 \cG^-(x)&=& \dsp -\frac{1}{k+2}\int d^2z 
 \,\left\{ 2(\bar{j}\Phi_1)\partial_x\eta 
 +\partial_x (\bar{j}\Phi_1) \eta \right\} \\
 \cT(x)&=& \dsp \{ Q_s,\, \sqrt{k}\cG^-(x)\}   \\
      &=& \dsp -\frac{1}{k+2} \int d^2z  \,
    \left\{(\bar{j}\Phi_1)(\partial_x^2 \tilde{j}
   + 4\partial_x\eta\partial_x\xi) \right. \\
  & & \dsp \hspace{1in} +  \left.
   \partial_x(\bar{j}\Phi_1) (\frac{1}{2}\partial_x \tilde{j}
   +3\partial_x \eta\xi+2\eta\partial_x\xi)
   +\partial_x^2(\bar{j}\Phi_1)\eta\xi  \right\} \\
   \cJ(x)&=& \dsp  -\frac{1}{k+2}  \int d^2z\, 
  \left\{(\bar{j}\Phi_1)\left(J_{\cN/U(1)}-\partial \rho
   +(k-1)\sqrt{\frac{2}{k}}K+2\partial_x\eta\xi\right)
   +\partial_x(\bar{j}\Phi_1)\eta\xi  \right\} 
\end{array} 
\label{stTCA2}
\end{equation}
In these expressions one should regard the products of operators 
inserted at the same points on world-sheet
as the standard normal product defined by
\begin{equation}
 A(z)B(z) \df \oint_z dw \, \frac{A(w)B(z)}{w-z}.
\label{np}
\end{equation}

Now, why can we claim that they are the correct currents in the  space-time 
conformal algebra?

First of all, we can derive the Ward identities of these currents
$\{\cT(x), \, \cG^{\pm}(x),\, \cJ(x)\}$ 
by means of the methods outlined 
in \cite{KS} {\em without using the free field approximations\/}.
We only have to evaluate the correlators having the forms;
$
\left\langle \partial_{\bar{x}}\cA(x)\, \cA '(y)\, \cO \ldots
\right\rangle ,
$
where $\cA(x)$, $\cA'(y)$ are one of $\{\cT(x), \, \cG^{\pm}(x),\,
\cJ(x)\}$, and $\cO \ldots$ means the suitable primary operators 
in space-time CFT.
In these calculations
we can make use of the formulas \eqn{jj}, \eqn{Phi 1 OPE}, \eqn{xi eta}
and among other things\footnote
     {This formula \eqn{Phi Phi} is justified only by semiclassical
      calculations in \cite{KS}. But, we here assume the validity 
     of this identity in quantum level.}
\begin{equation}
\Phi_1(x,\bar{x};z,\bar{z})
\Phi_h(y,\bar{y};z,\bar{z}) = \de^2(x-y)\Phi_h(y,\bar{y};z,\bar{z}) .
\label{Phi Phi}
\end{equation}
The following identities are also useful;
\begin{eqnarray}
\Phi_1 &=&\frac{1}{\pi}\partial_{\bar{x}}\La ,   \\
\bar{j}\Phi_1 &=& \frac{k+2}{\pi}\partial_{\bar{z}}\La , \\
\partial_z \Phi_1 
&=& \frac{1}{k+2}\partial_x(j \Phi_1) ,
\end{eqnarray}
where 
\begin{equation}
\La(x,\bar{x};z,\bar{z}) 
= \frac{1}{x-\gamma(z)} 
\frac{|x-\gamma(z)|^2 e^{\sqrt{\frac{2}{k}}\varphi(z)}}
{|x-\gamma(z)|^2 e^{\sqrt{\frac{2}{k}}\varphi(z)}+1} ,
\end{equation}
as well as  the next elementary identity of delta function;
\begin{equation}
 x^m \partial_x^n \delta(x)
  =\left\{
\begin{array}{ll} \dsp
 (-1)^m \frac{n!}{(n-m)!} 
\partial_x^{n-m}\delta(x) & (m \leq n) \\
  0 & (m>n) .
\end{array}\right. 
\end{equation}
We shall not present here the explicit forms of the Ward identities
among the currents $\{\cT(x), \, \cG^{\pm}(x),\, \cJ(x)\}$
\eqn{stTCA2}, but the important point is as follows:
These Ward identities are nicely
interpreted as the OPE's among them,
and we can prove that they truly generate  
a topological conformal algebra with 
\begin{equation}
\begin{array}{lll}
 \hat{c}&=& 2kI, \\
 I&=&\dsp \frac{1}{(k+2)^2}\int d^2z \, 
j(x;z)\bar{j}(\bar{x};\bar{z})\Phi_1(x,\bar{x};z,\bar{z}) .
\end{array} 
\label{I}
\end{equation}
Here $I$ is actually a constant on the boundary
$\partial_x I = \partial_{\bar{x}} I=0$, and under the long string 
configuration near boundary this constant reduces to
$\dsp \lim_{\varphi\rightarrow +\infty} I= \oint \gamma^{-1}d\gamma$.
(For the more detailed arguments, especially with respect to the
contribution from the short string sector, refer to \cite{DORT,KS}.)

Secondly, we can show that the currents
\eqn{stTCA2} reduce to the free field
realizations \eqn{stTCA1} in the near boundary limit
$\varphi \rightarrow +\infty$.
Under this limit we have 
$\dsp \La \sim \frac{1}{x-\gamma}$, and can recover the results in the  free
field approximations \eqn{stTCA1};
$\dsp \lim_{\varphi\rightarrow \infty}\,\oint dx x^{n+1}\cT(x)= \cL_n$,
$\dsp \lim_{\varphi\rightarrow \infty}\,\oint dx x^{n}\cG^+(x)= \cG^+_n$,
$\dsp \lim_{\varphi\rightarrow \infty}\,\oint dx x^{n+1}\cG^-(x)= \cG^-_n$,
and $\dsp \lim_{\varphi\rightarrow \infty}\,\oint dx x^{n}\cJ(x)= \cJ_n$.

Here we comment on a non-trivial point to take the limit 
$\varphi \rightarrow +\infty$. The operator products in the expressions
\eqn{stTCA2} are defined by \eqn{np}, but those for the near boundary 
results \eqn{stTCA1} stand for the normal ordering with respect to the mode
expansions of free fields. These two do not coincide in general.
Especially, 
(in the calculation of 
$\dsp \lim_{\varphi\rightarrow \infty}\,\oint dx x^{n}\cG^+(x)) $
it is important to remark the next identity under the limit 
$\varphi \rightarrow +\infty$;
\begin{equation}
  \partial_x j(x;z)\partial_x \Phi_1(x,\bar{x};z,\bar{z})
  = :\partial_x j(x;z)\partial_x \Phi_1(x,\bar{x};z,\bar{z}):
    +\frac{2}{k+2}\partial_x^2:(j(x;z)\Phi_1(x,\bar{x};z,\bar{z})):,
\end{equation}
where L.H.S means the normal product defined by \eqn{np} and 
the symbol ``$:~~:$'' in R.H.S means 
the normal ordered product of free fields.

There is an important consistency check for the space-time currents
\eqn{stTCA2}. We must show that $\{\cG^{\pm}(x),\, \cT(x),\, \cJ(x) \}$
are actually ``holomorphic'', that is,
they have no singular OPE's  with their anti-holomorphic counterparts 
$\{\bar{\cG}^{\pm}(\bar{y}),\, \bar{\cT}(\bar{y}),\, \bar{\cJ}(\bar{y})
\}$. It is not a trivial statement in contrast to the case of free
fields \eqn{stTCA1}, since the expressions of \eqn{stTCA2} 
include the terms such as $\sim \partial_x^*(\bar{j}\Phi_1)$
which may have singular OPE's with the anti-holomorphic currents.
For this purpose let us  first confirm the next identity;
\begin{equation}
\{\bar{Q}_s,\, \cA(x)\}=0,~~~(\forall \cA(x) 
\in \{\cG^{\pm}(x),\, \cT(x),\, \cJ(x)\}).
\label{Qs bar}
\end{equation}
We start with the next simple identity;
\begin{equation}
\lb \bar{Q}_s,~ \partial_x^n(\bar{j}\Phi_1) \rb=-\frac{k+2}{\sqrt{k}}
      \partial_{\bar{z}}\partial_x^n(\bar{\xi}\Phi_1).
\end{equation}
From this, we obtain 
\begin{equation}
\left\lb \bar{Q}_s,~ \int d^2z \, \partial_x^n(\bar{j}\Phi_1)
    F(j,\eta,\xi,\ldots) \right\rb = \int d^2z \, \partial_{\bar{z}}(*)=0,
\end{equation}
for an arbitrary polynomial $F$ of only  the holomorphic 
operators $j(x;z)$, $\eta(x;z)$, $\xi(x;z)$, $\ldots$, which proves 
the above identity \eqn{Qs bar}.

Nextly, it is not so hard  to prove 
\begin{equation}
\partial_{x}\bar{\cG}^-(\bar{x})\cA(y) \sim 0,~
\partial_{x}\bar{\cJ}(\bar{x})\cA(y) \sim 0,~~~
(\forall \cA(y) 
\in \{\cG^{\pm}(y),\, \cT(y),\, \cJ(y)\}) .
\label{bar G J}
\end{equation}
In the proof for them  we can make use of the next identitity
(and its cousins);
\begin{equation}
\oint_{\bar{w}} d\bar{z} \Phi_1(x,\bar{x};z,\bar{z})
(\bar{j}(\bar{y};\bar{w})\Phi_1(y,\bar{y};w,\bar{w}))
=- 2(\bar{x}-\bar{y})\delta^2(x-y)
 -(\bar{x}-\bar{y})^2 \partial_{\bar{x}}\delta^2(x-y)=0.
\end{equation}
Because of the above results \eqn{Qs bar}, \eqn{bar G J} and also 
the relations $\dsp \bar{\cG}^+= -\sqrt{k}\lb \bar{Q}_s ,~ \bar{\cJ} \rb$,
$\dsp \bar{\cT} = \sqrt{k}\{\bar{Q}_s,~ \bar{\cG}^-\}$,
we can also conclude that 
\begin{equation}
\partial_{x}\bar{\cG}^+(\bar{x})\cA(y) \sim 0,~
\partial_{x}\bar{\cT}(\bar{x})\cA(y) \sim 0,~~~
(\forall \cA(y) 
\in \{\cG^{\pm}(y),\, \cT(y),\, \cJ(y)\}) ,
\label{bar G J 2}
\end{equation}
which are no other than the claims we should confirm.

~

\section{Conclusions}

\cleqn
\hspace*{4.5mm}

In this paper we explored the topologically twisted fermionic string
theory on the general back-ground $AdS_3\times {\cal N}$
which is compatible with $N=2$ SUSY on world-sheet.
We summarized  the  structure of the world-sheet topological theory,
and showed that the dual space-time (boundary) theory should become
also a topological conformal field theory.
Our main result is the construction of space-time conformal algebra
from the world-sheet picture. Firstly we considered the near
boundary approximation, which is described by the Wakimoto free field
representation, and secondly we took  the more rigid formulation 
which is still valid far from the boundary according to that given in
\cite{KS}.  The space-time algebra we derived  became  a twisted $N=2$
superconformal algebra with $\dsp \hat{c}=2k\oint\,\gamma^{-1}d\gamma$.
Although this $\hat{c}$ is equal to that of the twisted version of
the space-time $N=2$ algebra given in \cite{GRBL},
these two are not the same. The generators of the latter are not 
BRST-invariant in the topological string theory, and so cannot be
regarded as the symmetry of the space-time topological theory
which is dual to the world-sheet topological theory. 
Nevertheless it might be interesting to study further the relation between
them. For example, one of the interesting questions is
whether or not there exists a good modification,
such as a similarity transformation, from the former to the latter.

As we emphasized, the topological string has only 
the world-sheet fermions with integral spins, and hence we cannot have 
spin fields. These field contents on world-sheet are analogus to that given
in \cite{ito}, which is based on the free field realizations of 
affine Lie superalgebra and is deeply connected  with the recent 
``hybrid formalism'' \cite{hybrid} for the $AdS_3\times S^3$ superstring 
(in the special case with $B_{NS}\neq 0$, $B_{RR}=0$). 
Especially, it may be worthwhile to point out that 
the complex scalar $\phi$ in the topological gravity sector \eqn{TG}
has the same world-sheet property 
as that of the ``ghost field'' in \cite{hybrid}.
It might be interesting to discuss in detail the relationship between 
the topologically twisted theory and the framework of hybrid formalism. 

Analysis on physical spectrum in our topological string theory
is significant.
As several authors pointed out \cite{SW,BDM,Pet} (see also 
\cite{old,teschner}),  
there are still several open problems
and pathologies  in (fermionic) string theory on $AdS_3$, in particular
with respect to the analyses on physical spectra. 
Fortunately, in the toplogically twisted theory 
it seems that the spectrum becomes rather simple and 
we need not face such pathologies.
This is because {\em only\/} the 1/2 BPS states can be physical in the 
topological theory. However, there may be a few subtleties  
in this spectrum, for example, about the field identification rules of 
chiral primary fields. We will study the spectra in detail for
some concrete examples in future work \cite{sugawara}.

~


\section*{Acknoledgement}

I would like to thank T. Eguchi, K. Ito, K. Sugiyama, Y. Satoh, and 
S. Yamaguchi for valuable discussions and several useful comments. 
I am also grateful to the organizers of Summer Institute '99 held at
Yamanashi, Japan, for their kind hospitality. 
Part of this work was done during my partitipation to there.

This work is supported in part  by 
Grant-in-Aid for Encouragement of Young Scientists ($\sharp 11740142$) 
and also by Grant-in-Aid for Scientific Research on Priority Area 
($\sharp 707$) ``Supersymmetry and Unified Theory
of Elementary Particles", 
both from Japan Ministry of Education, Science, Sports and Culture.


\newpage
\appendix
\section{Topological Conformal Algebra}

\cleqn
\hspace*{4.5mm}

We here present our convention of the topological conformal algebra
$\{L_n,\, G^{\pm}_n,\,J_n \}$ (or $\dsp T(x)=\sum_n\,
\frac{L_n}{x^{n+2}}$, $\dsp G^+(x)=\sum_n\,\frac{G^+_n}{x^{n+1}}$,
$\dsp G^-(x)=\sum_n\,\frac{G^-_n}{x^{n+2}}$, 
$\dsp J(x)=\sum_n\,\frac{J_n}{x^{n+1}}$)
which are defined from the 
$N=2$ suprconformal algebra (in Ramond sector)
by twisting with respect to the $U(1)_R$-current \cite{EY};  
\begin{equation}
L_n \df L_n^{N=2} -\frac{1}{2}(n+1)J_n ,
\end{equation}
\begin{eqnarray}
\lb L_m,~ L_n \rb &=& (m-n)L_{m+n} , \nonumber\\
\lb J_m,~ J_n \rb&=&\hat{c}m\delta_{m+n,0}, \nonumber\\
\{G_m^+,~G_n^-\}&=&L_{m+n}+mJ_{m+n}+\frac{\hat{c}}{2}m(m-1)\delta_{m+n,0},
\nonumber\\
\{G_m^{\pm},~ G_n^{\pm}\}&=&0 , \nonumber\\
\lb L_m,~G^{\pm}_n \rb &=& \{\frac{(1\mp 1)}{2}m-n\}G^{\pm}_{m+n} , \nonumber\\
\lb L_m,~J_n\rb &=& -nJ_{m+n}-\frac{\hat{c}}{2}m(m+1)\delta_{m+n,0},
\nonumber\\
\lb J_m,~ G^{\pm}_n\rb &=& \pm G^{\pm}_n ,
\end{eqnarray}
or equivalently,
\begin{eqnarray}
T(x)T(y) &\sim& \frac{2}{(x-y)^2}T(y)+\frac{1}{x-y}\partial_xT(y) \nonumber\\
J(x)J(y) &\sim& \frac{\hat{c}}{(x-y)^2} \nonumber\\
G^+(x)G^-(y)&\sim& \frac{\hat{c}}{(x-y)^3}+\frac{1}{(x-y)^2}T(y)
 +\frac{1}{x-y}J(y) \nonumber\\
G^{\pm}(x)G^{\pm}(y)&\sim& 0 \nonumber\\
T(x)G^{\pm}(y)&\sim&\frac{\frac{3}{2}\mp \frac{1}{2}}{(x-y)^2}G^{\pm}(y)
    +\frac{1}{x-y}\partial_y G^{\pm}(y) \nonumber\\
T(x)J(y)&\sim&-\frac{\hat{c}}{(x-y)^3}+\frac{1}{(x-y)^2}J(y)
+\frac{1}{x-y}\partial_y J(y) \nonumber\\
J(x)G^{\pm}(y)&\sim& \pm \frac{1}{x-y}G^{\pm}(y) .
\end{eqnarray}
This algebra is completely chracterized by only one parameter
$\hat{c}\, (\dsp \equiv \frac{c^{N=2}}{3})$.


\end{document}